\documentclass [12pt]{iopart}
\usepackage{iopams}
\usepackage{color,graphicx}
\newcommand{\OpL}{\hat{\mathcal{L}}}
\begin{document}
\title{Analytical theory of Doppler reflectometry in slab plasma model}
\author{E Z Gusakov and A V Surkov}
\address{Ioffe Institute, Politekhnicheskaya 26,
194021 St. Petersburg, Russia} \eads{
\mailto{a.surkov@mail.ioffe.ru}}
\begin{abstract}
Doppler reflectometry is considered in slab plasma model in the
frameworks of analytical theory. The diagnostics locality is
analyzed for both regimes: linear and nonlinear in turbulence
amplitude. The toroidal antenna focusing of probing beam to the
cut-off is proposed and discussed as a method to increase
diagnostics spatial resolution. It is shown that even in the case
of nonlinear regime of multiple scattering, the
 diagnostics can be used for an estimation (with certain accuracy) of plasma poloidal
 rotation profile.\end{abstract}
\pacs{52.70.Gw, 52.35.Hr, 52.35.Ra}
\section{Introduction}
\begin{figure}
\begin{center}
\includegraphics[height=0.2\textheight]{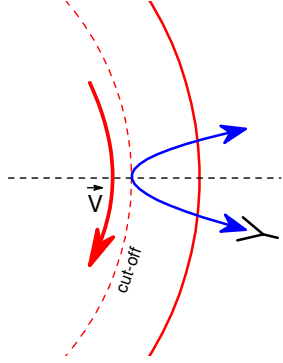}
\end{center}
\caption{\label{fig:Scheme} Diagnostics scheme. }
\end{figure}

One of widespread methods used nowadays for plasma rotation
velocity measurements is Doppler
reflectometry~\cite{Zou,Bulanin00,Hirsch01}. This technique
provides measuring fluctuations propagation poloidal velocity
which is often shown to be dominated by plasma poloidal rotation
velocity~\cite{Hirsch01}. The method is based on plasma probing
with a microwave beam which is tilted in respect to plasma density
gradient (see \fref{fig:Scheme}). A back-scattered signal with
frequency differing from the probing one is registered by a nearby
standing or the same antenna. The information on plasma poloidal
rotation is obtained in this technique from the frequency shift of
the backscattering (BS) spectrum which is supposed to originate
from the Doppler effect due to the fluctuation rotation.

Analytical theory of Doppler reflectometry was developed in recent
papers~\cite{We1,We2,We3}, using analytical approach in slab
plasma model, which is reliable for elongated plasma of large
tokamaks. The linear case of probing wave single-scattering is
considered there as well as the nonlinear process of the signal
formation due to multi-scattering effect, which is essential for
long probing ray trajectory, typical for large fusion devices. The
diagnostics spatial and wavenumber resolution is determined and
means to increase the method locality are discussed.

The present paper is devoted mainly to two following topics. First
of all, we modify the linear theory of Doppler reflectometry
taking into account possible antenna focusing in toroidal
direction, which allows the diagnostics spatial resolution to be
enhanced without deteriorating the poloidal wavenumber
selectivity, which takes place in case of poloidal focusing,
discussed in~\cite{We1}. Secondly, we compare the diagnostics
locality in linear and non-linear regime and dwell upon
experimental evidences, allowing us to distinguish these two
cases.
\section{Toroidal focusing in linear theory of Doppler reflectometry}
In our consideration we follow our paper~\cite{We1}, taking into
account possible antenna focusing in toroidal direction, which was
not considered there. The study is made in the frameworks of
geometrical optics (or WKB) approach, and the reader is referred
to~\cite{We1} for more accurate procedure, applied to the cut-off
vicinity, where WKB approximation is not valid.

We consider normalized antenna electric field in the following
form
\[
\vec E_a(\vec r)=\vec e_z \int\limits_{-\infty}^{+\infty}
\frac{\rmd k_y\rmd k_z}{(2\pi)^2} W(x,k_y,k_z) f(k_y,k_z)
\rme^{\rmi k_yy+\rmi k_zz}
\]
where $y$, $z$ axes denote poloidal and toroidal directions.
Factor $f(k_y,k_z)$ takes into account the antenna pattern
describing antenna radiation in vacuum
\[
f(k_y,k_z)=\sqrt{\frac{c}{8\pi }} \int\limits_{-\infty}^{+\infty}
\rmd y\,\rmd z E_0(x=0,y,z) \rme^{-\rmi k_yy-\rmi k_zz}
\]
where $E_0$ is vacuum antenna field, differing from $E_a$ by the
absence of the reflected wave contribution. We consider tilted
gaussian antenna pattern
\begin{equation}
f(k_y,k_z)=2\sqrt{\pi\rho_y\rho_z}\exp\left\{-\frac12\left[\rho_y^2\left(k_y-\mathcal{K}\right)^2+\left(\rho_z^2-\frac{\rmi
c\mathcal{R}}\omega\right)k_z^2\right]\right\}\label{eq:gauss}
\end{equation}
where a possibility to provide antenna focusing in toroidal
direction is taken into account. Corresponding parameter
$\mathcal{R}$ in case of
\begin{equation}
\frac{c\mathcal{R}}{\omega}\gg\rho_z^2\label{eq:condR}
\end{equation}
has a meaning of a wavefront curvature radius at the antenna.
In~\eref{eq:gauss} $\mathcal{K}$ corresponds to the antenna tilt
($\mathcal{K}=\omega/c\sin\vartheta$, where $\vartheta$ denotes
tilt angle in respect of the density gradient).

 According
to~\cite{Tyntarev}  radial distribution of  ordinary wave electric
field in WKB-approximation has the following form:
\begin{eqnarray*}
\fl W(x,k_y,k_z)=4\sqrt{\frac{2\pi \omega }{c^2 k_x(x,k_y,k_z)} }
\exp\left[\rmi\int_0^{x_c(k_y,k_z)} k_x(x',k_y,k_z)\,\rmd x'
-\frac{\rmi
\pi }{4} \right]  \\
\lo{\times} \cos\left[ \frac{\pi }{4} -\int_x^{x_c(k_y,k_z)}
k_x(x',k_y,k_z)\,\rmd x'\right]
\end{eqnarray*}
where  $k_x^2(x,k_y,k_z)=k^2(x)-k_y^2-k_z^2=[\omega ^2-\omega
_{pe}^2(x)]/c^2 -k_y^2-k_z^2$, the turning point $x_c(k_y,k_z)$ is
determined by the equation
\[
k_x\left[x_c(k_y,k_z),k_y,k_z\right]=0
\]
and $x=0$ corresponds to the plasma border.

The scattering signal according to reciprocity
theorem~\cite{Ginzburg, PiliyaPopov} can be written as
\[
A_s(\omega _s)=\frac{\rmi e^2}{4 m_e\omega }\sqrt{P_i}
\int_{-\infty}^{+\infty} \delta n_{\Omega}(\vec r) E_a^2(\vec
r)\,\rmd\vec r
\]
where $P_i$ is the probing wave power and $\delta n_{\Omega}(\vec
r)$ is the density fluctuation with frequency~$\Omega$.

Using the same procedure as described in~\cite{We1} we consider
the turbulence to be slightly inhomogeneous along $x$ direction
and rotating with plasma in the poloidal direction, so that the
density fluctuation correlation function takes the form
\begin{eqnarray}
\fl\left\langle\delta n(x,y,t_1)\delta
n(x',y',t_2)\right\rangle=\delta n^2\left(\frac{x+x'}{2}\right)
 \int_{-\infty}^{+\infty}\frac{\rmd\varkappa\,\rmd q\,\rmd\Omega }{(2\pi)^3}
\left| \tilde
n\left(\varkappa,q,\Omega\right)\right|^2\nonumber\\\lo{\times}
\exp\left[ \rmi\varkappa (x-x')+\rmi
q(y-y')-\rmi\Omega(t_1-t_2)-\rmi
qv\left(\frac{x+x'}{2}\right)(t_1-t_2)\right]\label{eq:inhom}
\end{eqnarray}
where $v(x)$ is the radial distribution of the plasma poloidal
velocity. This allows us to obtain a spectral power density of the
received signal in the following form~\cite{We1}
\begin{equation}
 p(\omega _s)=\langle
|A_s|^2\rangle=P_i\int_{-\infty}^{+\infty} \rmd x \,\delta
n^2(x)S(x)\label{eq:lin}
\end{equation}
The scattering efficiency $S(x)$ can be shown to consist of
backscattering (BS) and forward scattering (FS) contributions
\begin{equation*}
 S(x)=\frac{1}{2}
\left(\frac{e^2}{m_e c^2 }\right)^2 \int_{-\infty}^{+\infty}
\frac{\rmd q}{2\pi }\left[ S_{BS}(x,q)+S_{FS}(x,q)\right]
\end{equation*}
where
\begin{eqnarray}
\fl S_{BS}(x,q)=
\frac{\left|f\left(-q/2,0\right)\right|^4}{k_x^2\left(x,\mathcal{K},0\right)}
\sum_{m=\pm 1}\left|\tilde
n\left[2mk_x\left(x,\mathcal{K},0\right),q,\Omega-qv(x)\right]\right|^2\nonumber\\\lo{\times}
{\left\{\rho_y^4+\frac{c^2}{\omega^2}\left[\Lambda_0+m\Lambda(x)\right]^2\right\}^{-1/2}\left\{\rho_z^4+
\frac{c^2}{\omega^2}\left[\Lambda_0-\mathcal{R}+m\Lambda(x)\right]^2\right\}^{-1/2}}
\label{eq:SBS}
\end{eqnarray}
Here $m=\pm1$ corresponds to BS after and before the cut-off in
respect of the probing ray propagation.  The FS efficiency takes
the form
\begin{eqnarray}
\fl S_{FS}(x,q)=
2\left\{\rho_y^4+\frac{c^2}{\omega^2}\Lambda_0^2\right\}^{-1/2}
\left\{\rho_z^4+\frac{c^2}{\omega^2}(\Lambda_0-\mathcal{R})^2\right\}^{-1/2}
\left|f\left(-\frac{q}2,0\right)\right|^4
\nonumber\\
\lo{\times} \exp\left\{-\frac12
\left[\frac{\rho_yq\Lambda(x)}{\Lambda_0}\right]^2\right\}
k_x^{-2}\left(x,\mathcal{K},0\right)\left|\tilde n\left[\frac{q^2
\Lambda(x)}{2k(x)\Lambda_0} ,q,\Omega-qv(x)\right]\right|^2
\label{eq:SFS}
\end{eqnarray}
where
\[
\Lambda(x)=\frac{\omega }{c}\int_x^{x_c(\mathcal{K},0)} \frac{\rmd
x'}{k_x(x',\mathcal{K},0)},\quad \Lambda_0\equiv \Lambda(0)
\]

We consider expressions~(\ref{eq:SBS}),~(\ref{eq:SFS}) from the
diagnostics locality point of view. First of all, the locality is
determined by reversed square of the radial
wavenumber~$k_x^{-2}\left(x,\mathcal{K},0\right)$. This factor,
corresponding to WKB-behavior of antenna electric field,
underlines the cut-off vicinity, but for unfavorable density
profiles, for example, linear or bent down ones, it does not
suppress enough plasma periphery contribution.  For the BS the
second factor is fluctuations spectral density~$\left|\tilde
n\left[\pm2k_x\left(x,\mathcal{K},0\right),q,\Omega\right]\right|^2$.
Due to the dominance of long scales in the turbulence spectrum
this factor underlines the cut-off vicinity, where
$k_x\left(x,\mathcal{K},0\right)$ is small. For FS contribution
the signal suppression due to the acquisition by antenna pattern
periphery, described by the factor
\[
\exp\left\{-\frac12
\left[\frac{\rho_yq\Lambda(x)}{\Lambda_0}\right]^2\right\}
\]
 plays an analogical role, it underlines the cut-off
vicinity, where $\Lambda(x)$ is small.

Additional localization for BS contribution can be provided by
antenna focusing to the cut-off, which occurs at
$\mathcal{R}=\Lambda_0$. If the beam is narrow enough in toroidal
direction, so that condition~(\ref{eq:condR}) is satisfied for
$\mathcal{R}=\Lambda_0$,
  the factor
\[
\left\{\rho_z^4+
\frac{c^2}{\omega^2}\left[\Lambda_0-\mathcal{R}+m\Lambda(x)\right]^2\right\}^{-1/2}=
\left\{\rho_z^4+
\left[\frac{c\Lambda(x)}{\omega}\right]^2\right\}^{-1/2}
\]
  is large in the cut-off
vicinity. It should be noted that, the focusing in the poloidal
direction can not give us such an effect, due to the fact, that we
can not provide narrow antenna beam in poloidal direction without
deteriorating the diagnostics poloidal wavenumber selectivity,
which is described by $\left|f\left(-q/2,0\right)\right|^4$.

Considering toroidal focusing influence on the FS efficiency, it
should be noted that
  the focusing increases the amplitude of FS
signal~(\ref{eq:SFS}), but does not improve its locality.

\begin{figure}[t]
\begin{center}
\includegraphics[width=\textwidth]{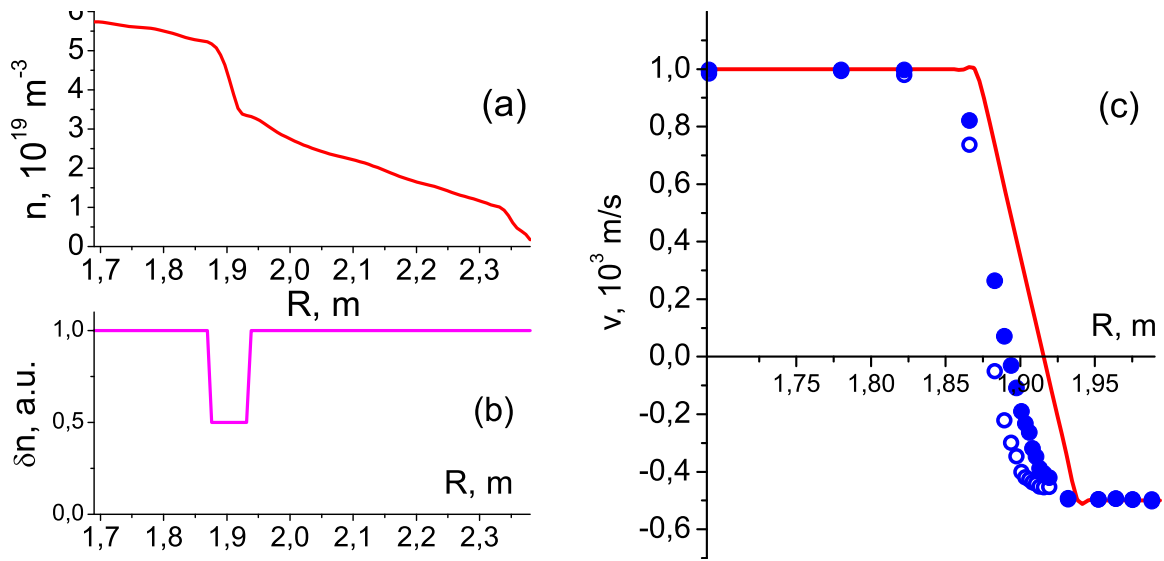}
\end{center}
\caption{\label{fig:Focus} Antenna focusing influence. (a)~DIII-D
density profile~\cite{Doyle}. (b)~Turbulence amplitude assumed.
(c)~Poloidal velocity profile ({\color{red}\full}), and velocity
estimated using Doppler reflectometry signal frequency spectrum
shift: {\color{blue}\textbullet}---using antenna focusing,
{\color{blue}$\circ$}---without focusing.}
\end{figure}
The influence of the factors discussed on the spectrum of the
registered signal is illustrated in the modelling, described
in~\cite{We1}, and the results  can be found below
in~\sref{sec:Disc} (see~\fref{fig:Nlin1} and~\fref{fig:Nlin2}) in
comparison with spectra, modelled in nonlinear diagnostics regime.
Now let us dwell upon the spectrum modelling, illustrating the
antenna toroidal focusing influence. We consider the density
profile of \mbox{DIII-D} tokamak plasma with internal transport
barrier (\fref{fig:Focus}(a))~\cite{Doyle}. The probing is
performed at different frequencies and therefore with different
cut-off positions. Here we take into account the distance between
antenna and the plasma, which was assumed to be equal~$1$~m, and
suppose that focusing is performed into narrow in the toroidal
direction spot ($\rho_z\sim1$~cm) to provide
condition~(\ref{eq:condR}) to be satisfied for
$\mathcal{R}=\Lambda_0$. Besides that we take into account the
turbulence suppression in the barrier region
(see~\fref{fig:Focus}(b)).

Despite the fact that density profile in the barrier region is
favorable for the diagnostics~\cite{We1}, antenna focusing makes
the spectrum shift more adequate to the behavior of plasma
velocity in the cut-off.
\section{Noninear theory of Doppler reflectometry}
In the section we review briefly the nonlinear analytical theory
of Doppler reflectometry, which is considered in details
in~\cite{We2,We3}. In case of long enough trajectory length and
sufficient turbulence amplitude, when the following
criterion~\cite{GusPopov}
\begin{equation}
\gamma\equiv\frac{\omega_i^2}{c^2}\,\left(\frac{\delta
n}{n_c}\right)^2x_c\ell_{cx}\ln\frac{x_c}{\ell_{cx}}\gtrsim1\label{eq:crit}
\end{equation}
 is satisfied, where $\omega_i$ is the probing frequency,
$\delta n/n_c$ is the turbulence amplitude, normalized to the
density in the cut-off, $x_c$ is the distance to the cut-off, and
$\ell_{cx}$ is the turbulence radial correlation length,  we can
neglect the BS during the wave propagation and consider multiple
FS of the probing wave only. The density fluctuations in this case
can be taken into account as a phase modulation during the probing
wave propagation to the cut-off and backward. It can be shown that
condition~(\ref{eq:crit}) holds true in large plasma devices even
at small density perturbation level $\delta n/n_c\lesssim10^{-2}$.

The wave electric field is determined by Helmholtz equation
\begin{equation*}
\Delta E+[k^2(x)+\delta k^2(x,y,t)]E=0
\end{equation*}
 where
\[
\delta k^2(x,y,t)=-\frac{\omega_i^2}{c^2}\frac{\delta
n(x,y,t)}{n_c}
\]
 is the fluctuation of the wavenumber (here it is
given for ordinary wave, but it can be easily written for
extraordinary wave too). The electric field can be represented in
the following form
\begin{equation}
E( l,y)=\int_{-\infty}^{+\infty} G[
l,y|0,y_0;t]E_a^{(i)}(y_0)\,\rmd y_0\label{eq:E}
\end{equation}
where $l$ is a coordinate along the ray trajectory and
\[
\fl G\left[l,y|0,y_0;t\right]=\sqrt{\frac{\omega_i}{2\pi c
l}}\exp\left\{ \frac{\rmi\omega_i}{2c}\left[\frac{(y-y_0)^2}{ l}
-\frac1{n_c}\int_0^{ l}n\left[x( l'),y^{(0)}( l'),t\right]\,\rmd
l'\right] -\frac{\rmi\pi}{4}\right\}
\]
\Eref{eq:E} describes the transportation of initial condition from
the plasma border, where probing antenna is situated, inside the
plasma along the ray trajectory. A function $G$ contains the
turbulence phase shift in question, which is determined by the
density fluctuations and represents multiple FS effect.

According to the reciprocity theorem~\cite{Ginzburg,PiliyaPopov},
the registered signal is determined as an integral over all plasma
border of the wave reflected by the cut-off
\begin{equation}
A_s=\frac{c}{16\pi}\int_{-\infty}^{+\infty}\rmd y E( 2\Lambda_0,y)
E_a^{(r)}(y)\label{eq:nlin}
\end{equation}
with a weight function, determined by the electric field of the
acquisition antenna~$E_a^{(r)}(y)$, if we consider it as probing
one. Actually, due to narrow in the wavenumber space antenna
pattern, the main component in this signal is formed due to the
multiple FS, which changes essentially a poloidal wavenumber of
the probing wave, meanwhile the radial wavenumber changes the sign
due to reflection off the cut-off.

To obtain the spectrum of the registered signal and at the same
time to analyze the diagnostics locality, as above we consider
inhomogeneous turbulence, poloidally rotating with a
plasma~(\ref{eq:inhom}). Averaging  $|A_s|^2$~(\ref{eq:nlin}) we
obtain the spectrum in question~\cite{We2,We3}
\begin{equation}
S(\omega)\propto \exp\left\{
   -\frac12\frac{
     \left[
       \omega-\omega_i+2\mathcal{K}\left(\rho^{-2}+\OpL q^2\right)^{-1}\OpL q^2v\right]^2
   }{
     \OpL \left( \Omega^2+q^2v^2 \right)-\left( \rho^{-2}+\OpL q^2 \right)^{-1}
     \left( \OpL q^2v \right)^2
   }
\right\}\label{eq:spectr}
\end{equation}
Here an operator $\OpL$
has the meaning of the integration over distance from the plasma
border to the cut-off, with the averaging over the turbulence
spectrum
\begin{eqnarray}
\fl\OpL\xi\simeq\frac{\omega_i^2}{c^2n_c^2} \int_0^{x_c}\rmd
x\,\delta n^2(x)\int_{-\infty}^{+\infty}\frac{\rmd\varkappa\,\rmd
q\,\rmd\Omega}{(2\pi)^3}\, \left|\tilde
n\left(\varkappa,q,\Omega\right)\right|^2\xi\nonumber\\\lo{\times}
\left\{
    \begin{array}{lll}
       \omega_i^2/c^2\,\delta(\varkappa)\,k^{-2}(x),\, & x_c-x>\ell_{cx}/4\\
       4L, & x_c-x\le\ell_{cx}/4\\
       \end{array}
\right.\label{eq:L}
\end{eqnarray}
where $L=[\rmd\ln n_e(x)/\rmd x|_{x=x_c}]^{-1}$ is the density
variation scale in the cut-off. The integration in~\eref{eq:L} is
performed with a weight function, proportional to the
inhomogeneous turbulence amplitude, and the factor, underlying the
cut-off vicinity.

To analyze this expression for signal spectrum  we consider simple
case of homogeneous plasma poloidal rotation $v(x)=v$, which gives
$\hat\mathcal{L}q^2 v=v\hat\mathcal{L}q^2$. In case of strong
nonlinear regime, when antenna beam divergence is completely
determined by the turbulence ($\rho^2\hat\mathcal{L}q^2 \gg1$) it
can be seen that
 spectrum frequency shift is determined by traditional (linear)
Doppler effect
\begin{equation}
\omega_{max}=\omega_i-2\mathcal{K}v\label{eq:shift1}
\end{equation}
On the contrary, the  spectrum broadening
\begin{equation*}
\Delta\omega=\sqrt{\hat\mathcal{L}\Omega^2}
\end{equation*}
 is
strongly influenced in nonlinear case by the turbulence amplitude
and differs from linear one
\begin{equation*}
\Delta\omega_{lin}=\left[\int_{-\infty}^{+\infty}\frac{\rmd\varkappa\,\rmd
q\,\rmd\Omega}{(2\pi)^2}\,
 \left|n\left(\varkappa,q,\Omega\right)\right|^2\Omega^2\right]^{1/2}
\end{equation*}
by the factor, which  can be estimated as
$\Delta\omega/\Delta\omega_{lin}\sim\gamma$, where $\gamma$ is
determined by~\eref{eq:crit}. Thus in nonlinear case the frequency
spectrum width can be substantially larger than in linear
situation, when the factor $\gamma$ is similar or less than $1$.

In case of inhomogeneous plasma poloidal rotation the spectrum
frequency shift is actually determined by the specifically
averaged rotation velocity
\begin{equation}
\omega_{max}=\omega_i-\frac{2\mathcal{K}\hat\mathcal{L}
q^2v}{\rho^{-2}+\hat\mathcal{L}q^2 }\label{eq:shift2}
\end{equation}
In this case the frequency spectrum shift can be produced by the
region with high amplitude of the turbulence as well as by the
region with high poloidal velocity.

The frequency spectrum broadening, which in case of homogeneous
plasma poloidal rotation is caused by intrinsic frequency spectrum
of fluctuations, is influenced here by additional factor
associated with poloidal rotation inhomogeneity.
\[
\Delta\omega=\left[\hat\mathcal{L}\left(\Omega^2+q^2v^2
\right)-\frac{\left(\hat\mathcal{L}
q^2v\right)^2}{\rho^{-2}+\hat\mathcal{L}q^2 }\right]^{1/2}
\]

\section{Discussion}
\label{sec:Disc} At first, we consider the spectrum modelling,
which was carried on according to the results of linear and
nonlinear consideration of the Doppler reflectometry.

In calculation the following assumptions are made:
\begin{figure}
\begin{center}
\includegraphics[width=\textwidth]{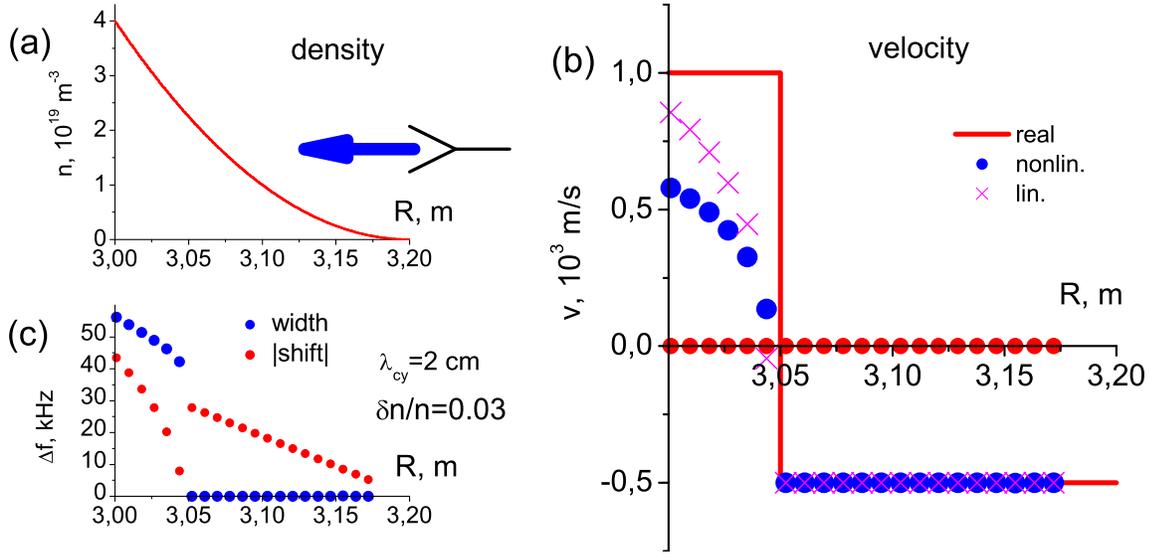}
\end{center}
\caption{\label{fig:Nlin1} The signal spectrum evolution.
(a)~Assumed density profile. $R$ denotes  the major radius.
(b)~Assumed poloidal velocity profile ({\color{red}\full}),
cut-off positions ({\color{red}\textbullet}) and measured poloidal
velocity profile in nonlinear ({\color{blue}\textbullet}) and
linear ({\color{magenta}$\times$}) diagnostics regimes.
(c)~Absolute value of frequency shift ({\color{red}\textbullet})
and width ({\color{blue}\textbullet}) of signal spectrum related
to different probing frequencies (plotted via corresponding
cut-off position).}
\end{figure}
\begin{enumerate}
\item Geometrical parameters taken correspond to Tore Supra
experiments~\cite{Zou}: $\omega_i/c\sim 12\mbox{ cm}^{-1}$,
$\rho\sim 14\mbox{~cm}$, $\vartheta\sim 11.5^\circ$ distance to
the cut-off $L\sim 20\mbox{~cm}$. \item We consider the same
antenna for the probing and reception. The probing is performed at
different frequencies and therefore with different cut-off
positions.  \item For the sake of simplicity we suppose the
turbulence level to be uniform ($\tilde n=0.03$) and its
wavenumber spectra to be gaussian.  The fluctuations are believed
to be low-frequency to neglect the spectrum width in case of
homogeneous poloidal velocity profile.
\end{enumerate}
First of all we  consider plasma density
profile~(\fref{fig:Nlin1}(a)) similar to observed in Tore
Supra~\cite{Clairet} and step-like plasma poloidal velocity
distribution~(\fref{fig:Nlin1}(b)). The registered signal
frequency shift~(\fref{fig:Nlin1}(c)) is calculated
using~\eref{eq:shift2}, but, as it is usually done in experiment
results interpretation, the measured poloidal velocity
profile~(\fref{fig:Nlin1}(b)) is deduced from frequency spectrum
shift using equation~\eref{eq:shift1} for traditional (linear)
Doppler effect in assumption that the registered signal spectrum
shift corresponds to velocity in the cut-off.

It can be seen that the value of poloidal velocity measured in
such a way coincides with assumed one in case of homogeneous
poloidal velocity distribution and differs from it when the
cut-off is situated in the region of variable velocity. In this
case large contribution to the $\hat\mathcal{L}q^2v$ value is made
by far from the cut-off regions due to unfavorable bent-down
density profile, leading to the obscuration of the velocity in the
cut-off. Thus such an interpretation of the diagnostics results
gives the value of the poloidal velocity averaged in a specific
way over plasma volume.

The frequency spectrum shift is compared in~\fref{fig:Nlin1}(c)
with spectrum width, calculated for $\ell_{cy}\sim2$~cm. It can be
seen that spectrum width can be larger then frequency spectrum
shift, which is typical for Doppler reflectometry experimental
results. Besides, \fref{fig:Nlin1}(c) demonstrates the spectrum
width behavior which was described above: the spectrum is not
broadened when probing wave propagates only in the region with
homogeneous poloidal velocity (here we neglected the broadening,
which arises due to intrinsic frequency spectrum of the
turbulence) and broadens, when the cut-off, which is the bound of
propagation region, crosses the point, where the velocity changes.

In addition, in~\fref{fig:Nlin1}(b) we compare the diagnostics
results in linear and nonlinear regimes. In the both cases the
plasma density profile is unfavorable for the diagnostics spatial
resolution, but one can see that nonlinear regime is more
problematic from the results interpretation point of view.

The second example is to emphasize the importance of plasma
density profile for frequency spectrum formation.
\begin{figure}
\begin{center}
\includegraphics[width=\textwidth]{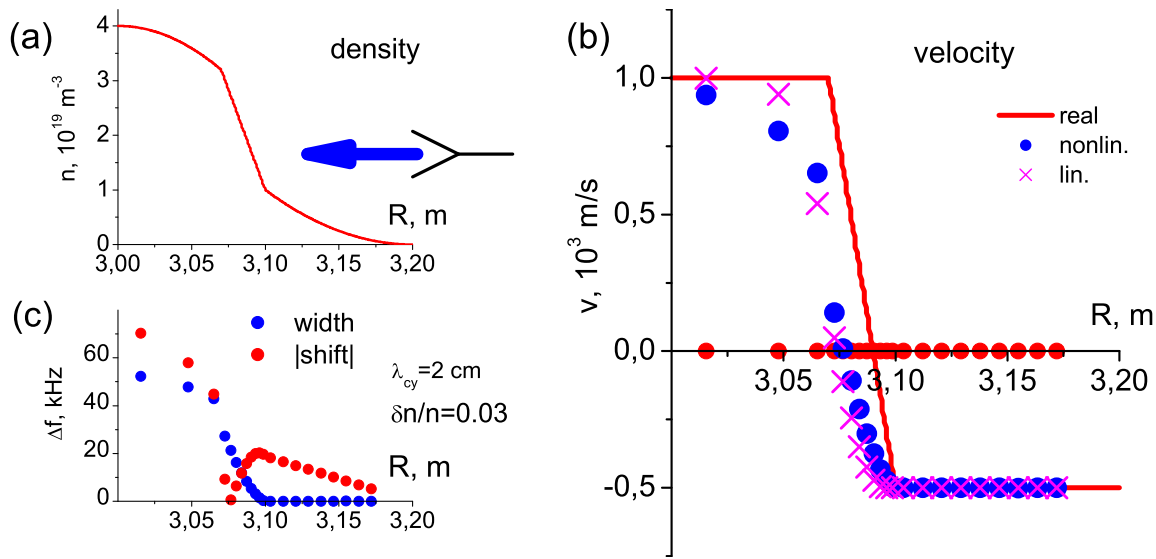}
\end{center}
\caption{\label{fig:Nlin2} The signal spectrum evolution.
(a)~Assumed density profile. $R$ denotes  the major radius.
(b)~Assumed poloidal velocity profile ({\color{red}\full}),
cut-off positions ({\color{red}\textbullet}) and measured poloidal
velocity profile in nonlinear ({\color{blue}\textbullet}) and
linear ({\color{magenta}$\times$}) diagnostics regimes.
(c)~Absolute value of frequency shift ({\color{red}\textbullet})
and width ({\color{blue}\textbullet}) of signal spectrum related
to different probing frequencies (plotted via corresponding
cut-off position).}
\end{figure}
We consider plasma density profile~(\fref{fig:Nlin2}(a)) bent down
in plasma periphery and bent up in the core. The poloidal velocity
profile has high gradient in the `barrier'
region~(\fref{fig:Nlin2}(b)). It is easy to see that bent-up
density profile underlines the cut-off contribution and improves
the locality of the method.  Also this effect is essential for
linear Doppler reflectometry~\cite{We1}. It leads to the fact that
in case of favorable density profile the poloidal velocity profile
measured by Doppler reflectometry corresponds to the certain
extent to the real one.

Let us discuss the situation when the nonlinear theory developed
should be applied. The nonlinear regime of Doppler reflectometry
diagnostics is considered in the present paper, when multiple FS
influence is essential for registered signal spectrum formation.
For this situation to take place, two important criteria should be
satisfied. At first, the turbulence amplitude is to be large
enough to provide essential turbulent phase shift during the wave
propagation, which is described by criterion~(\ref{eq:crit}).

The second condition leading to the necessity of nonlinear theory
application to the diagnostics results is that small-angle
scattering contribution is substantial in the received signal.
This means that small-angle scattering signal amplitude should be
comparable or larger than BS  signal, formed by linear mechanism.
Multiple small-angle scattering contribution~\eref{eq:nlin} can be
evaluated as
\[
\left\langle\left|A_s\right|^2\right\rangle\sim\frac{P_i}{4}\frac{\omega_i\ell_{cy}^2}{cx_c\gamma}
\exp\left\{-\frac{2\mathcal{K}^2\ell_{cy}^2}{\gamma^2}\right\}
\]
The BS signal amplitude~(\ref{eq:lin}) can be estimated as
\[
\left\langle\left|A_s^{lin}\right|^2\right\rangle\sim\frac{P_i}{2\sqrt{2\pi}}\gamma\frac{\omega_i\rho}{cx_c}\left|n(-2\mathcal{K})\right|^2
\]
Here $\left|n(q)\right|^2$ is the fluctuation poloidal wavenumber
spectrum. The ratio of the signal amplitudes
\[
\alpha\equiv\frac{\left\langle\left|A_s\right|^2\right\rangle}{\left\langle\left|A_s^{lin}\right|^2\right\rangle}\sim\frac{\ell_{cy}^2}{\rho\gamma^2}\frac{\exp\left\{-2\mathcal{K}^2\ell_{cy}^2/\gamma^2\right\}}{\left|n(-2\mathcal{K})\right|^2}
\]
For example, we consider turbulence spectral density
$\left|n(q)\right|^2=4\pi\ell_{cy}\left[1+q^2\ell_{cy}^2\right]^{-3/2}$,
where $\ell_{cy}\sim2$~cm, probing frequency
$f=\omega_i/2\pi=60$~GHz, tilt angle $\theta=30^\circ$,
$\rho=2$~cm. Then if $\gamma\sim20$ the small-angle scattering
contribution larger than BS one: $\alpha\sim1.5$.

\begin{figure}
\begin{center}
\includegraphics[height=0.2\textheight]{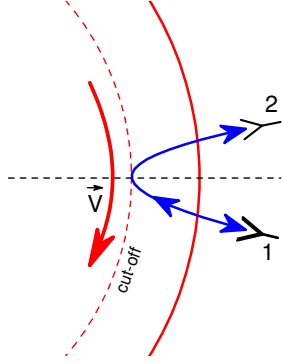}
\end{center}
\caption{\label{fig:Bi} Diagnostics scheme. 1---emitting and
receiving antenna, 2---additional receiving antenna.}
\end{figure}
To conclude, we consider the experimental evidences of linear or
nonlinear regime.  A reliable criteria, which is actually very
difficult to realize in experiment, is the spectrum of the passed,
reflected off the cut-off signal. If the probing line can be
distinguished  in the spectrum of the signal, registered by
antenna $2$ in~\fref{fig:Bi}, we deal with linear regime of the
scattering. In other case, when the probing line can not be
observed in the broadened spectrum, the diagnostics works in
nonlinear regime.

Another way to recognize the scattering regime is to compare BS
signal spectrum width with one, provided by the antenna pattern
due to the Doppler effect. In case of more broadened spectrum we
deal with nonlinear regime of scattering.

\begin{figure}
\begin{center}
\includegraphics[height=0.2\textheight]{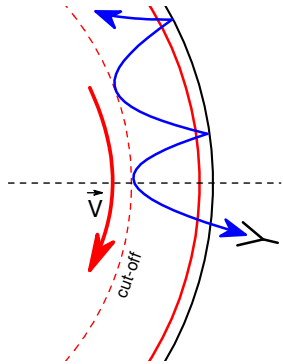}
\end{center}
\caption{\label{fig:Multi} Multi-scattering in small toroidal
device.}
\end{figure}
The criteria of nonlinear theory applicability discussed can be
usually satisfied  in large toroidal devices. But we should note
that the nonlinear regime can be realized in small tokamak due to
possible reflection of the probing signal off the chamber wall
(this effect for FT-2 tokamak was investigated experimentally
in~\cite{Gurch}). It leads to the effective increase of the ray
trajectory, which can provide the satisfaction of the
criterion~(\ref{eq:crit}). The signal spectrum of the signal can
be estimated in this case using the results of our
consideration~(\ref{eq:spectr}), but the operator
$\OpL$~(\ref{eq:L}) should be multiplied by the quantity of the
reflections  off the plasma and wall.

\section{Conclusion}
In the present paper the Doppler reflectometry is considered in
slab plasma model in the frameworks of analytical theory. The
diagnostics locality is analyzed for linear and nonlinear regime.
The toroidal beam focusing to the cut-off is proposed and
discussed as a method to increase diagnostics spatial resolution.

In nonlinear diagnostics regime frequency spectrum shift and width
of registered backscattered signal is analyzed in dependence on
plasma density profile, turbulence spatial distribution and
spectrum and plasma poloidal velocity profile. It is demonstrated
that the frequency shift is not influenced by turbulence absolute
amplitude and gives an information on poloidal velocity averaged
over the vicinity of the cut-off, the size of which depends on the
density profile and turbulence distribution.

Thus, even in the complicated situation of multi-scattering
dominance Doppler reflectometry technique is proved to be able to
give realistic information on plasma rotation. The spatial
resolution of the diagnostics, however, suffers from transition to
this nonlinear regime of scattering. The consideration presented
allows the spatial resolution of the method to be analyzed for
real experimental conditions, and thus the diagnostics results to
be adequately interpreted.
\ack The support of RFBR grants 02-02-17589, 04-02-16534, INTAS
grant 01-2056 and NWO-RFBR grant 047.016.015 is acknowledged.
A.V.~Surkov is thankful to the``Dynasty'' foundation for
supporting his research.
\end{document}